\documentclass[aps,prl,reprint,groupedaddress]{revtex4-1}
\usepackage{mathptmx,siunitx,graphicx,amssymb}
\usepackage[version=3]{mhchem}
\usepackage{multibib}

\DeclareSIUnit{\tex}{tex}
\DeclareSIUnit{\sq}{sq}

\newcites{supp}{Supp References}

\begin{document}
\title{Bio-Inspired Aggregation Control of Carbon Nanotubes for Ultra-Strong Composites}
\author{Yue Han,$^{1,3}$ Xiaohua Zhang,$^2$ Xueping Yu,$^2$ Jingna Zhao,$^2$ Shan Li,$^{1,3}$ Feng Liu,$^1$ Peng Gao,$^4$
Yongyi Zhang,$^2$ Tong Zhao,$^1$ Qingwen Li$^2$}
\affiliation{\footnotesize $^1$Laboratory of Advanced Polymeric Materials, Institute of Chemistry, Chinese Academy of
Sciences, Zhongguancun North First Street 2, Beijing 100190, China\\ $^2$Key Laboratory of Nano-Devices and
Applications, Suzhou Institute of Nano-Tech and Nano-Bionics, Chinese Academy of Sciences, Ruoshui Road 398, Suzhou
215123, China\\$^3$University of Chinese Academy of Sciences, Yuquan Road 19, Beijing 100049, China\\$^4$Suzhou Creative
Nano Carbon Co.\ Ltd., Ruoshui Road 398, Suzhou 215123, China\\E-mail: xhzhang2009@sinano.ac.cn; tzhao@iccas.ac.cn;
qwli2007@sinano.ac.cn}

\begin{abstract}
High performance nanocomposites require well dispersion and high alignment of the nanometer-sized components, at a high
mass or volume fraction as well. However, the road towards such composite structure is severely hindered due to the easy
aggregation of these nanometer-sized components. Here we demonstrate a big step to approach the ideal composite
structure for carbon nanotube (CNT) where all the CNTs were highly packed, aligned, and unaggregated, with the
impregnated polymers acting as interfacial adhesions and mortars to build up the composite structure. The strategy was
based on a bio-inspired aggregation control to limit the CNT aggregation to be sub 20--50 nm, a dimension determined by
the CNT growth. After being stretched with full structural relaxation in a multi-step way, the CNT/polymer
(bismaleimide) composite yielded super-high tensile strengths up to 6.27--6.94 GPa, more than 100\% higher than those of
carbon fiber/epoxy composites, and toughnesses up to 117--192 MPa. We anticipate that the present study can be
generalized for developing multifunctional and smart nanocomposites where all the surfaces of nanometer-sized components
can take part in shear transfer of mechanical, thermal, and electrical signals.
\end{abstract}

\maketitle

A composite material is typically made up of two or more constituent materials with significantly different physical or
chemical properties, and is also named a nanocomposite when one of the constituents has one, two, or three dimensions of
less than 100 nm. To design the structure for high performance and multifunctional composites, nature has offered us
with scientific and technological clues from the formation of biological composites using common organic components via
the naturally mild approaches \cite{ha.tlb:2013}. For example, super-tough spider fibers are derived from desirable
organization of linear protein molecules \cite{giesa.t:2011}, strong hard nut skins are assembled from the mixture of
cellulose and lignin molecules \cite{preston.cm:1992}, and wear-resistant molluscan shells are a result of
biomineralization of calcium carbonates in a brick-and-mortar manner \cite{porter.sm:2007}. To make these natural
composites mechanically strong, a homogeneous distribution of the major components such as proteins, cellulose
molecules, and nanometer-sized crystals of carbonated calcium phosphates or calcium carbonates is a key structural
feature \cite{giesa.t:2011, cheng.qf:2014}. Their desired orientation along with other co-existing components also sheds
lights on the way to stronger man-made nanocomposites. This means, in order to fabricate high performance
nanocomposites, the fraction of nanometer-sized components should be as high as possible, while the other components
should act as interfacial adhesions and mortars to combine the major parts together. As a result, the interfacial
contacts or bondings can be maximized to allow full utilization of the unique properties of the nanometer-sized
components.

Owing to the superior mechanical properties of carbon nanotubes (CNTs), many composite structures have been proposed for
pursuing a wide range of industrial applications of CNT over the past two decades \cite{baughman.rh:2002,
coleman.jn:2006, moniruzzaman.m:2006, liu.lq:2011, kong.lr:2014}. As CNTs are difficult to be uniformly dispersed within
polymer matrix at a high mass fraction due to their strong tendency to agglomerate \cite{coleman.jn:20062,
fiedler.b:2006, xie.xl:2005, spitalsky.z:2010, rahmat.m:2011}, it is still a challenge to fabricate CNT composites that
mimic the natural ones. Fortunately, CNTs can be treated as linear macromolecules, and thus the processing on them can
be dealt with in a biomimic way. To mimic the formation process of biological composites, the preformed two-dimensional
(2D) CNT assemblies like sheets and films, whose thickness is within tens to hundreds of nanometer or over 1 \si{\micro
m}, are interesting candidates \cite{liu.lq:2011, zhang.m:2005}. By introducing thermosetting polymers like bismaleimide
(BMI) into these 2D assemblies, it has been possible to synthesize CNT composites at a high CNT mass fraction
\cite{cheng.qf:2009, cheng.qf:2010, liu.w:2011, di.jt:2012, wang.x:20131}, whose tensile strength was even higher than
that of T300 carbon fiber/epoxy composites (1.86 GPa) \cite{t300}. However, besides the high length-to-width aspect
ratio and high mass fraction, a set of structural parameters are still severely required, such as a high CNT packing
density and alignment, efficient matrix-to-CNT interfacial stress transfer, and, most importantly, the avoid of CNT
aggregation \cite{wagner.hd:2007}.

\begin{figure}[!t]
\centering
\includegraphics[width=.48\textwidth]{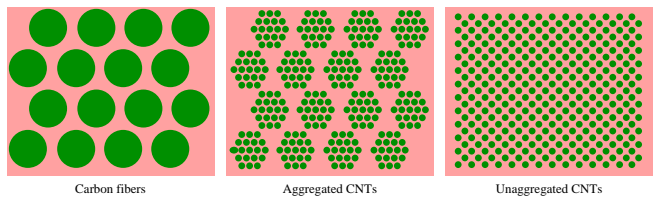}
\caption{\label{fig.schematic} Schematics of carbon fiber reinforced polymer, composite structure with aggregated CNTs,
and the ideal structure containing unaggregated CNTs, respectively.}
\end{figure}

The necessity of aggregation control can be demonstrated by a comparison between the structures of carbon fiber
reinforced polymer and CNT composites containing aggregated and unaggregated CNTs, as schematically shown in Figure
\ref{fig.schematic}. The most important advantage of CNTs is the large contact area between CNTs and matrix, similar to
the natural composite structures. When solid carbon fibers are replaced by the aggregated CNTs, as commonly observed in
today's CNT composites \cite{cheng.qf:2009, cheng.qf:2010, liu.w:2011, di.jt:2012, wang.x:20131}, the interfacial
contact area becomes much larger. The composites based on aggregated CNTs have exhibited tensile strengths ranging from
2.08 to 3.8 GPa \cite{cheng.qf:2009, cheng.qf:2010, wang.x:20131}. However, the aggregation phase might act as weak
parts in transferring external loads and thus hinders the further reinforcement. In an ideal structure where all the
interfaces can play roles in shear load transfer, the nanometer-sized components should be uniformly distributed within
the matrix and there should be no aggregation for either the matrix or the nanometer-sized structures.

Here we report a big step to approach such ideal structure where the composite structure contained highly aligned and
unaggregated CNT bundles. By learning the formation process of biological composites, polymers were impregnated into CNT
networks to obtain the uniform dispersion of the CNTs among the polymer matrix. As the CNTs were well covered by the
polymers, sufficient stretching exercises were performed to improve the CNT alignment with maintaining the CNT
aggregation level below 20--50 nm, and to increase the mass density as well. The new CNT composites exhibited ultra-high
and stable tensile strengths up to 6.27--6.94 GPa and toughnesses up to 117--192 MPa, corresponding to the energies
absorbed before rupturing of 75--124 \si{\J\per\g} by considering the mass density of $\sim$1.55 \si{\g\per\cubic\cm}.
Such tensile strengths have been more than 100\% higher than those of carbon fiber/epoxy composites. The processing
method is supposed to be generalized for developing multifunctional and smart nanocomposites where all the surfaces of
nanometer-sized components can take part in shear transfer of mechanical, thermal, and electrical signals.


{\bf Entangled CNT network.} The CNT aggregation arises from van der Waals (vdW) attraction and can be enhanced in wet
environment. The situation becomes very severe in the layer-by-layer stacking of aligned CNT sheets with the aid of
solution spray to obtain high performance CNT films \cite{liu.w:2011, di.jt:2012, wang.x:20131}, where the CNTs (or more
commonly, small-sized CNT bundles) usually aggregate first into large-sized bundles and then are surrounded by polymer
matrix, as discussed below. Instead, preformed CNT networks can be the optimal raw materials \cite{cheng.qf:2009,
cheng.qf:2010}.

The networked CNTs were synthesized by using an injection chemical vapor deposition (CVD) method \cite{li.yl:2004},
where a mist of ethanol, ferrocene, and thiophene was injected into a heated gas flow reactor (see Supplementary
Information). The grown CNTs cross-linked with each other, formed a sock-like aerogel in the gas flow, and were blown
out with the carrier gas, a mixture of Ar and \ce{H2}. By continuously winding the CNT aerogel on a roller with the aid
of ethanol densification, CNT films with a thickness of 10--30 \si{\micro m} were obtained.

\begin{figure}[!t]
\centering
\includegraphics[width=.48\textwidth]{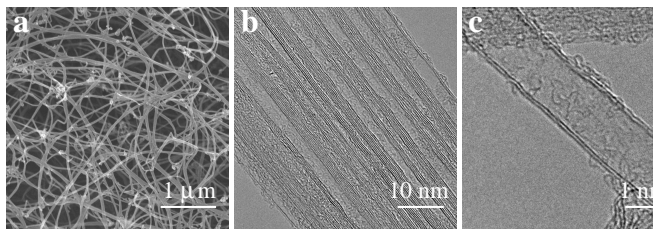}
\caption{\label{fig.entanglement} Assembly structure of as-produced CNT films. (a) CNT bundles contacted with each other
and formed a network. (b) The bundle size was $\sim$50 nm in width. (c) Double-walled CNTs were the major growth
output.}
\end{figure}

Scanning electron microscopy (SEM) and transmission electron microscopy (TEM) have shown that the basic structural units
of the as-produced CNT films were small-sized bundles (Figure \ref{fig.entanglement}a), with a diameter of 40--50 nm and
containing about 50 CNTs (Figure \ref{fig.entanglement}b). The CNTs in a bundle usually grew out of the same iron
catalyst nanoparticle and thus became always bundled during the growth. Under the gas flow, the bundles contacted with
each other and finally formed an entangled assembly structure. The CNTs were mainly double-walled and had a diameter of
1--2 nm (Figure \ref{fig.entanglement}c), and were confirmed with Raman spectroscopy (see Supplementary Information,
Figure \ref{fig.raman}). The as-produced CNT films had a high level of crystallinity as reflected by the high G to
D-band Raman intensity ratio, and the CNT mass fraction was over more than 90\% (see Supplementary Information, Figure
\ref{fig.raman}). Furthermore, nitrogen adsorption/desorption measurement revealed a specific surface area of 119
\si{\m\square\per\g}.

{\bf Impregnation without introducing aggregation.} It is very important to find that the CNT entanglement was not
altered after liquid densification. After further being densified with acetone, the pore sizes of CNT films decreased
from $>$500 nm (Figure \ref{fig.entanglement}a) to $\sim$100--200 nm (see Supplementary Information, Figure
\ref{fig.densified}), while the feature of random distribution and unaggregation did not change. Such process is
reminiscent of the formation process of biological composites, where the matrix co-exists with and disperses the major
components as they are simultaneously grown from stem cells. Their fractions are always optimized during the growth to
allow the maximized interfacial stress transfer \cite{wagner.hd:2007}. Thus the processing sequence was modified to
introduce impregnation of polymer solution prior to any other processing that might damage the network, to avoid CNT
aggregation. (Notice that, besides the severe CNT aggregation \cite{liu.w:2011, di.jt:2012, wang.x:20131}, the
pre-aligned CNT sheets drawn out of CNT arrays \cite{zhang.m:2004} are not favorite also because that it is difficult to
wet them as they are mechanically very weak.)

\begin{figure*}[!t]
\centering
\includegraphics[width=0.65\textwidth]{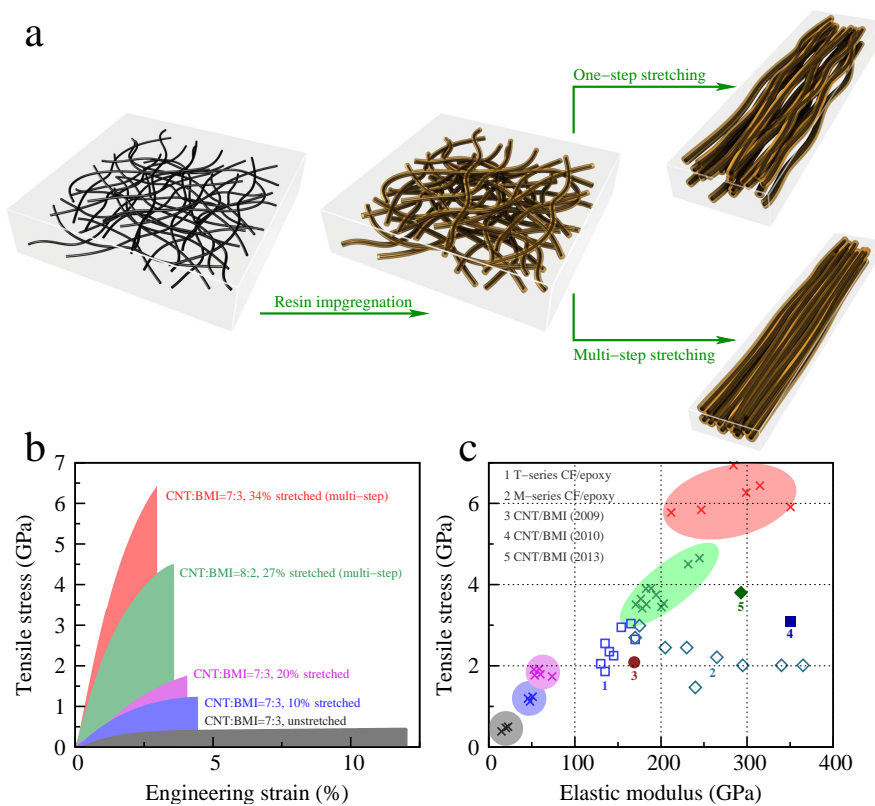}
\caption{\label{fig.tensile} Schematic of stretching methods and mechanical properties of CNT/BMI composite films. (a)
The polymer solution is impregnated into CNT films prior to the stretching process. The processing can be performed in
one-step and multi-step ways, which result in different levels of CNT alignment and densification. (b) Typical
stress-strain curves for CNT/BMI composite films prepared by different stretching methods. (c) Comparison of tensile
strength and modulus of different CNT/BMI composite films, labelled by $\times$ with the same color shown in (b), to the
T- and M-series carbon fiber/epoxy composites and recently reported high performance CNT/BMI composite films
\cite{cheng.qf:2009, cheng.qf:2010, wang.x:20131}.}
\end{figure*}

By appropriately using acetone as solvent to dissolve thermosetting polymers or their resins, like BMI resins, the
polymers can efficiently cover all the CNT surfaces. Excitingly, neither the CNTs nor the resins formed aggregated
phases; there did not exist a region filled with only CNTs or BMI resins above a size scale of 50--100 nm (see below the
detailed characterization). The entanglement played the key role, because the capillary force due to solvent evaporation
and the vdW interaction between CNT and resin could densify the assembly by drawing the CNTs closer, but these
interactions were not strong enough to break the network cross-links and thus to aggregate the CNTs.

{\bf High alignment and ultra-high strengths.} Stretching should be provided to re-assemble the network and align CNTs.
This requires the samples to possess high plasticity. The raw films could be stretched by 10--15\% in length and owing
to the improved alignment their tensile strength increased from 180--198 to 500--600 MPa (see Supplementary Information,
Figure \ref{fig.raw-tensile}). For the ``wet'' films where 1 wt\% BMI resin/acetone solutions were impregnated to reach
a CNT-to-resin mass ratio of 7:3 or 8:2, the unstretched films became more plastic and fractured above a strain of
20--25\%, corresponding to higher processability (see Supplementary Information, Figure \ref{fig.wet-tensile}). This
means that the impregnation prior to stretching also resulted in improved processability.

If hot-pressing was applied on the unstretched films to cure resins, the CNT/BMI films finally exhibited a tensile
strength just of 478--501 MPa and strain at break of 10--12.2\%. On the contrast, by first stretching the ``wet'' film
by 20\% and then curing the film, the tensile strength and strain at break became 1.74--1.92 GPa and 3.4--5.2\%,
respectively (see Supplementary Information, Figure \ref{fig.one-step}). 

Nevertheless, $\sim$2 GPa was not the up limit. By further modifying the stretching method to a multi-step way (Figure
\ref{fig.tensile}a), the ``wet'' films could be stretched by 27--34\%. The total stretching process was carried out in
multiple steps. In each step 3\% additional stretching according to the immediate film length was applied and then 5--10
minutes were used to relax the films. The total stretching magnitude, for example, was $1.03^8 - 1 = 0.267$ or
$1.03^{10} - 1 = 0.344$ for 8 or 10 steps, respectively. The multi-step method fully aligned the CNTs and improved
packing density during the hot-pressing (owing to the decreased level of CNT waviness and less unstretched network
connections). At this stage, the basic structural units (the small-sized CNT bundles) were well surrounded by the BMI
resin molecules and maintained unaggregated phases. After being cured, the CNT/BMI composite films stably exhibited an
extremely high tensile strength up to 4.5--6.94 GPa (Table \ref{tab.strength}, and also see Supplementary Information,
Figure \ref{fig.multi-step}), depending on the CNT-to-resin mass ratio and the total stretching magnitude. At the same
time, the elastic modulus was up to 232--315 GPa and the strain at break became 2.7--4.5\%. Figure \ref{fig.tensile}b
shows the typical stress-strain curves for various CNT/BMI composite films and Figure \ref{fig.tensile}c provides the
comparison with carbon fiber/epoxy composites.

\begin{table}
\caption{\label{tab.strength} Mechanical properties of CNT/BMI composite films.}
\centering
\begin{tabular}{ccccc}
\hline
    & & & Toughness & Strain at break \\
No. & Strength (GPa) & Modulus (GPa) & (MPa) & (\%)\\
\hline
\multicolumn{5}{l}{CNT-to-resin ratio 7:3, multi-step stretched by 34\%}\\
1 & 6.940 & 284.2 & 191.7 & 4.33 \\
2 & 6.438 & 314.9 & 114.6 & 2.97 \\
3 & 6.265 & 299.0 & 117.1 & 3.10 \\
4 & 5.907 & 350.6 & 82.1  & 2.34 \\
5 & 5.842 & 246.8 & 104.3 & 3.17 \\
6 & 5.773 & 211.9 & 163.2 & 4.49 \\
\hline
\multicolumn{5}{l}{CNT-to-resin ratio 7:3, multi-step stretched by 25\%}\\
1 & 6.309 & 148.7 & 197.9 & 5.41 \\
2 & 5.781 & 127.9 & 173.3 & 5.41 \\
3 & 5.130 & 168.0 & 187.4 & 5.47 \\
4 & 4.467 & 111.3 & 127.1 & 5.02 \\
5 & 4.266 & 146.5 & 123.8 & 4.51 \\
6 & 3.826 & 153.6 & 100.4 & 3.95 \\
\hline
\multicolumn{5}{l}{CNT-to-resin ratio 8:2, multi-step stretched by 27\%}\\
1 & 4.651 & 244.7 & 83.0 & 2.87 \\
2 & 4.505 & 231.5 & 105.9 & 3.58 \\
3 & 3.748 & 194.3 & 89.8 & 3.58 \\
4 & 3.646 & 176.6 & 92.9 & 3.84 \\
5 & 3.515 & 183.0 & 83.3 & 3.60 \\
6 & 3.506 & 170.9 & 78.3 & 3.38 \\
\hline
\end{tabular}
\end{table}

{\bf Effect of low-softening-point resins.} The high mechanical performance also came from the low-softening-point
($<$60 \si{\celsius}) BMI resins (l-BMI) which were traditional BMI monomers modified with diallyl bisphenol A (DBA)
\cite{li.zm:2001}. As a comparison, the same DBA modification was applied on BMI monomers with larger molecular weights
and higher molecular rigidity, to synthesize BMI resins with higher softening point of $>$80 \si{\celsius} (h-BMI). The
low softening point resulted in a soft and viscous state even at room temperature, like soft wax, and thus the CNT film
impregnated with BMI resins was called as a ``wet'' film. The less-``wet'' CNT/h-BMI films (CNT-to-resin mass ratio 7:3)
could be only stretched directly by 16\% while the ``wet'' CNT/l-BMI films (7:3) were by $>$20\% (see Supplementary
Information, Figure \ref{fig.BMI}). Of great importance, the ``wet'' feature also allowed structural relaxation as
sufficient as possible during the multi-step stretching process. Notice that, the 16\%-stretched CNT/h-BMI composite
films exhibited a tensile strength of $\sim$1.5 GPa, as comparable to the CNT/l-BMI composites (see Supplementary
Information, Figure \ref{fig.one-step}), indicating that the major difference of the resins was the plasticizing ability
rather than the strengthening ability.

\begin{figure}[!t]
\centering
\includegraphics[width=.48\textwidth]{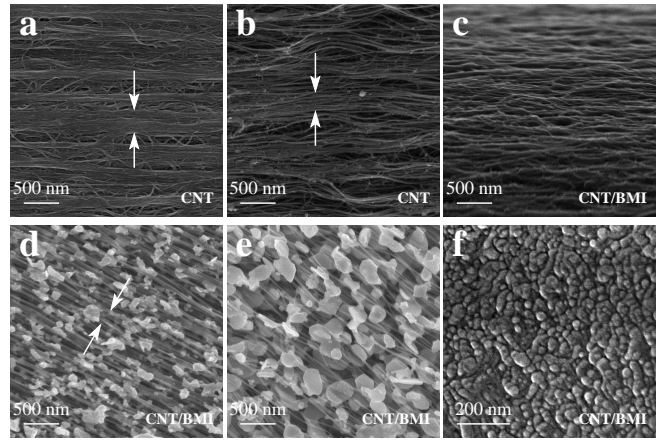}
\caption{\label{fig.sem} Comparison of CNT assembly structure for different films. (a) CNT aggregation in the
layer-by-layer stacking of array-drawn CNT sheets. (b) CNT aggregation in the stretched dry films composed by entangled
CNTs. (c) The small-sized CNT bundles did not aggregate but were surrounded by BMI polymers and uniformly distributed.
(d,e) 750 \si{\celsius} treated CNT/BMI composite films where the stretching of 21\% and 27\% was performed before and
after resin impregnation, respectively. (f) Cross section of the optimal CNT/BMI composite structure by using focused
ion beam treatment.}
\end{figure}

{\bf Structural characterization.} Based on these advantages, including the unaggregation and high alignment, we have
made a big step to realize the ideal composite structure. The tensile strength of $>$6 GPa is obviously much larger than
those of carbon fiber/epoxy composites (Figure \ref{fig.tensile}), in good agreement with their different composite
structures (Figure \ref{fig.schematic}). To show how much the present structure had approached the ideal one, comparison
was performed between the layer-by-layer stacked array-drawn CNT sheets, highly stretched film with entangled CNTs, and
the optimal and ultra-strong CNT/BMI composite films (Figure \ref{fig.sem}a-c). In the first two cases, CNT aggregation
was widely observed in a scale of hundreds of nanometer (Figure \ref{fig.sem}a,b), while the small-sized CNT bundles did
not aggregate but were surrounded and adhered with each other by BMI polymers in the optimal composite structure (Figure
\ref{fig.sem}c).

Thermal treatment at 750 \si{\celsius} in Ar for 1.5 h was performed on the CNT/BMI composite films to decompose BMI
polymers. After the decomposition, the remaining BMI polymers formed flake-like particles, and thus exposed the CNTs
which might maintain their aggregation level. Therefore this method can serve as evidences for CNT aggregation and
unaggregation. Two CNT/BMI composite films were tested, where the stretching process was performed before and after the
impregnation of BMI resins, respectively. As shown in Figure \ref{fig.sem}d and e, there was clearly no aggregation of
CNT bundles by using our new processing method.

It was possible to observe directly the cross section of the optimal structure by using focused ion beam treatment
(Figure \ref{fig.sem}f). Although it was difficult to distinguish individual CNTs, the small-sized CNT bundles were
pictured perpendicular to the cross section and polymer matrix surrounded all their surfaces. Such homogeneity maximized
interface contacts and thus provided the most efficient CNT-to-polymer stress transfer. By considering the fact that the
bundling within a dimension of 20--50 nm was only determined during the growth, there is still a last step to obtain the
ideal structure where individual CNTs are aligned, highly packed, and unaggregated.

Furthermore, it is no doubt that the stretching treatment improved the CNT alignment, nevertheless, quantitative
characterization of alignment is still of great interest. In the present study, the characterization was represented by
the Herman's orientation factor (HOF) which has been used to study the alignment level for CNT arrays \cite{xu.m:2012}.
HOF takes the value 1 for a system with full alignment and zero for completely nonoriented structures. The HOF of the
original film was only 0.209, well reflecting the random CNT distribution. After being stretched by about 20\% and 34\%,
the HOF increased remarkably up to 0.632 and 0.816, respectively. (The calculation method and the detailed results are
provided in Supplementary Information.)

{\bf Specific strengths.} Another way to describe the tensile property is specific strength, also known as the
strength-to-weight ratio. In this way it not necessary to know the film thickness. For the films multi-step stretched by
$\sim$25\% in length, the total mass for a 2 cm$\times$1 cm sample was 1.15 mg, corresponding to an area density of 0.58
\si{\micro\g\per\square\cm}. The fracture force per sample width was about 16.5 \si{\N\per\mm} in average (see
Supplementary Information, Figure \ref{fig.load}), and thus the specific strength (by dividing the force per width by
the area density) was $\sim$2.87 \si{\N\per\tex}. When the stretching magnitude was improved to 34\%, the fracture force
per width was $\sim$19.5 \si{\N\per\tex} (see Supplementary Information, Figure \ref{fig.load}), the area density
decreased slightly to 0.46 \si{\mg\per\square\cm}, and the specific strength was $\sim$4.24 \si{\N\per\tex}.

The volumetric mass density for the 34\% stretched film was measured to be $\sim$1.55 \si{\g\per\cubic\cm} according to
Archimedes' principle, where the film became suspended within a mixed solution containing dichloromethane (\ce{CH2Cl2})
and diiodomethane (\ce{CH2I2}) with a volume ratio of 13:2. Therefore, the engineering strength (product of specific
strength and volumetric density) should be about 6.57 GPa.

\begin{figure}[!t]
\centering
\includegraphics[width=.48\textwidth]{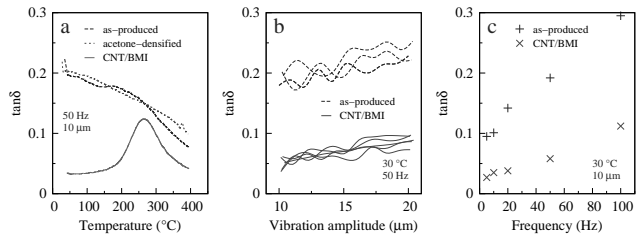}
\caption{\label{fig.damping} Loss factor of various films as functions of temperature (a), vibration amplitude (b), and
frequency (c).}
\end{figure}

{\bf Damping properties.} The entanglement of CNTs resulted in high damping performance for the as-produced dry films
(Figure \ref{fig.damping}). The loss factor $\tan\delta$ at 50 Hz was nearly 0.2 at room temperature, decreased
gradually to 0.1 as being heated up to 400 \si{\celsius} (Figure \ref{fig.damping}a), and linearly increased with
vibration amplitude (Figure \ref{fig.damping}b) and frequency (Figure \ref{fig.damping}c). The mechanism for the high
damping performance was suggested to be the sliding and de-bonding between CNT bundles.

For the composite structures, these energy-cost phenomenon nearly disappeared as the BMI polymers made all the bundles
adhered to each other and there no longer existed the so-called ``interfaces'' between different aggregation phases. At
room temperature, the loss factor was even smaller than 0.05 for low-frequency vibrations (Figure \ref{fig.damping}c).
However, due to the glass transition of polymer, the loss factor increased remarkably above 200 \si{\celsius}.

The high loss factor of the as-produced films (comparable to rubber) to makes it possible to develop new-type high
damping materials, based on the arrangement-induced viscoelastic behavior \cite{xu.m:2010}, while the CNT/BMI composite
films can be developed as superior structural materials to be used in aerospace, automotive, and other transportation
industries.

{\bf Electrical properties.} The ability to conduct electricity of a thin film is usually characterized by sheet
resistance or square resistance, in units of ``ohms per square''. The square resistance of the as-produced film was
1.194 \si{\ohm\per\sq}, by using the four-point probe method. After being impregnated with BMI resins, the resistance
decreased to 0.926 \si{\ohm\per\sq}, owing to the densification effect. After being stretched, the CNT network was
aligned and the connections between CNT bundles were separated by BMI resins. As a result, the resistance increased to
1.461 \si{\ohm\per\sq}. The curing process finally fixed the composite structure where neither CNTs nor polymers
aggregated, and the resistance further increased to 1.931 \si{\ohm\per\sq}. By considering the final film thickness of
$\sim$3 \si{\micro\m}, the electrical conductivity was $\sim1.7\times10^5$ S \si{\per\m}, about 0.3\% or 12\% of
copper's or stainless steel's electrical conductivity.

In summary, based on the unique properties of raw materials of CNTs and resins (entanglement, unaggregation, high
plasticity, and low softening point) and the multi-step stretching method, we have been able to obtain a magic composite
structure where neither CNTs nor polymers formed aggregated phases, a big step to approach the ideal composite structure
that can fully utilize all the CNT surfaces in load transferring. The highest tensile strength was up to 6.94 GPa (or
about 4.24 \si{\N\per\tex}), much higher than the strength of carbon fiber reinforced polymers. The CNT/BMI composite
films also exhibited high ability to conduct electricity. As in such composite structure nearly all the surfaces of
nanometer-sized components can be used, based on the bio-inspired aggregation control, we anticipate that the present
fabrication method can be generalized for developing multifunctional and smart nanocomposites.

{\bf Methods.} The CNTs were mainly double-walled and were synthesized with an injection chemical vapor deposition method
\cite{li.yl:2004}. The grown CNTs formed a sock-like aerogel and were winded on a roller with the aid of ethanol
densification to obtain 2D CNT films. The as-produced CNT films were impregnated by BMI resin/acetone solutions with
designed CNT-to-resin mass ratios. The optimal CNT-to-resin mass ratio was about 7:3. Then the resin-impregnated films
were stretched by more than 30\% in length, in a multi-step way where sufficient structural relaxation was allowed after
every step. The stretched films were cured according to the designed profile, namely, 140 \si{\celsius} for 0.5 h, 170
\si{\celsius} for 3 h, 220 \si{\celsius} for 2 h, and 250 \si{\celsius} for 3 h, with a pressure of 6--8 MPa.

Tensile tests were performed on an Instron 3365 Universal Test Machine (Instron Corp., Norwood, USA) at a strain rate of
0.5 \si{\milli\m\per\minute}. The film samples were cut into 2.5--3 cm$\times$0.5--2 mm pieces, and the gauge length was
larger than 10 mm. Some tensile tests to show the processability were also performed on the ``wet'' films with a larger
width of 5--10 mm.

Dynamic mechanical analysis was carried out with a Netzsch DMA 242E Analyzer (Netzsch-Ger\"{a}tebau GmbH, Selb,
Germany). Temperature-dependent loss factor ($\tan\delta$) was measured in the temperature range of $\sim$30--400
\si{\celsius} at a heating rate of 10 \si{\celsius} \si{\per\minute} and a vibration amplitude of 10 \si{\micro\m}.
Another scanning mode was performed where the vibration amplitude was tuned from 10 to 20 \si{\micro\m}, at room
temperature. The allowed vibration frequencies included 5, 10, 20, 50, and 100 Hz. The sample's gauge length was 6 mm,
corresponding to the dynamic vibration strain was 0.17\%--0.33\%.

{\bf Acknowledgements} This work was supported in part by the National Natural Science Foundation of China (21273269,
11302241, 21473238, 51473171), International Science \& Technology Cooperation Project of Jiangsu Province (BZ2011049),
and Suzhou Industrial Science and Technology Program (ZXG201416).

{\bf Author contributions} Y.H.\ fabricated the composite films and participated in the method design with X.Z.\ and
Q.L.; X.Y.\ and J.Z.\ repeated all the experiments that Y.H.\ had performed; Y.H., X.Y., J.Z., and S.L.\ performed
mechanical tests and structural characterizations; F.L.\ and T.Z.\ fabricated the resin molecules and assisted the
fabrication of composite films; P.G.\ and Y.Z.\ performed the injection chemical vapor deposition and provided the raw
CNT films; all authors provided information for manuscript writing and X.Z.\ wrote the manuscript with their helps.

\section*{Supplementary Information}

\setcounter{figure}{0}
\renewcommand{\thefigure}{S\arabic{figure}}

\subsection*{1. Preparation of entangled carbon nanotube films}

The CNTs were synthesized by using an injection chemical vapor deposition (CVD) method \citesupp{s-li.yl:2004}, where a
mist of ethanol, ferrocene (2 wt\%), and thiophene (1 vol\%) was injected at a rate of 20--30 ml/h into a heated gas
flow reactor (diameter 80 mm). A gas mixture of Ar (3500 sccm) and \ce{H2} (4250 sccm) were also injected into the
reactor tube as a carrier gas. The temperature in reaction region was set to 1300 \si{\celsius}. The grown CNTs formed a
sock-like aerogel in the gas flow and were blown out with the carrier gas. The CNT aerogel was winded on a roller with
the aid of liquid densification (ethanol was used here). By controlling the winding number, CNT films with a thickness
ranging from 10--30 \si{\micro\m} were finally obtained.

The basic structural units of the as-produced CNT films were small-sized CNT bundles, which usually had a diameter of
40--50 nm and contained about 50 CNTs. Once the ferrocene concentration or the injection rate is increased, larger-sized
bundles can be obtained. By controlling the growth parameters, such as the concentration of ferrocene and growth
temperature, it was possible to change the number of CNTs in a bundle and number of walls of individual CNT.

\begin{figure}[!t]
\centering
\includegraphics[width=.48\textwidth]{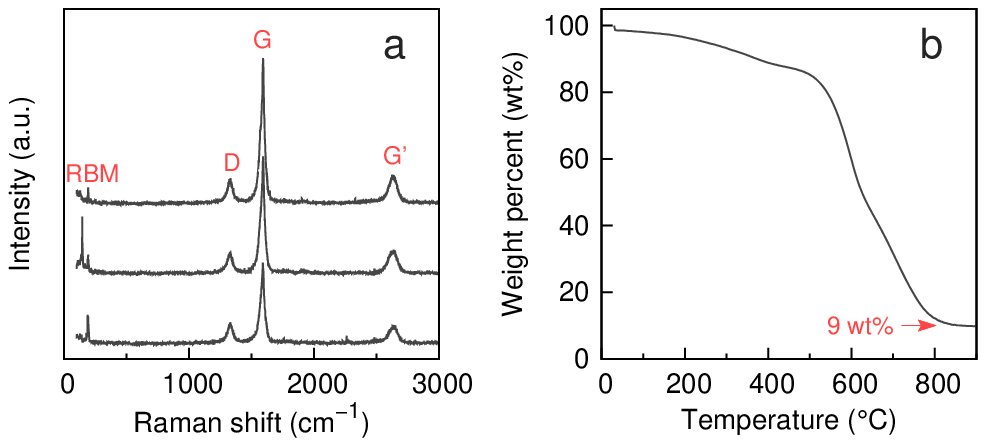}
\caption{\label{fig.raman} Characterization of as-produced CNT films. (a) Raman spectra obtained at different positions
of one film. (b) Thermal gravimetric analysis indicated that the CNT content was more than 91 wt\% by considering the
oxidation of iron.}
\end{figure}

Raman spectra of the CNTs showed a high G to D-band intensity ratio ($>$5), corresponding to a high level of
crystallinity (Figure \ref{fig.raman}a). The Raman signals of radial breathing modes (RBMs) also indicated that there
existed a large number of double-walled CNTs. Further, thermal gravimetric analysis showed that there was about 9 wt\%
mass left for the as-produced CNT films after being heated up to 800 \si{\celsius} in air (Figure \ref{fig.raman}b).

\subsection*{2. Entanglement after liquid treatment}

\begin{figure}[!t]
\centering
\includegraphics[width=.40\textwidth]{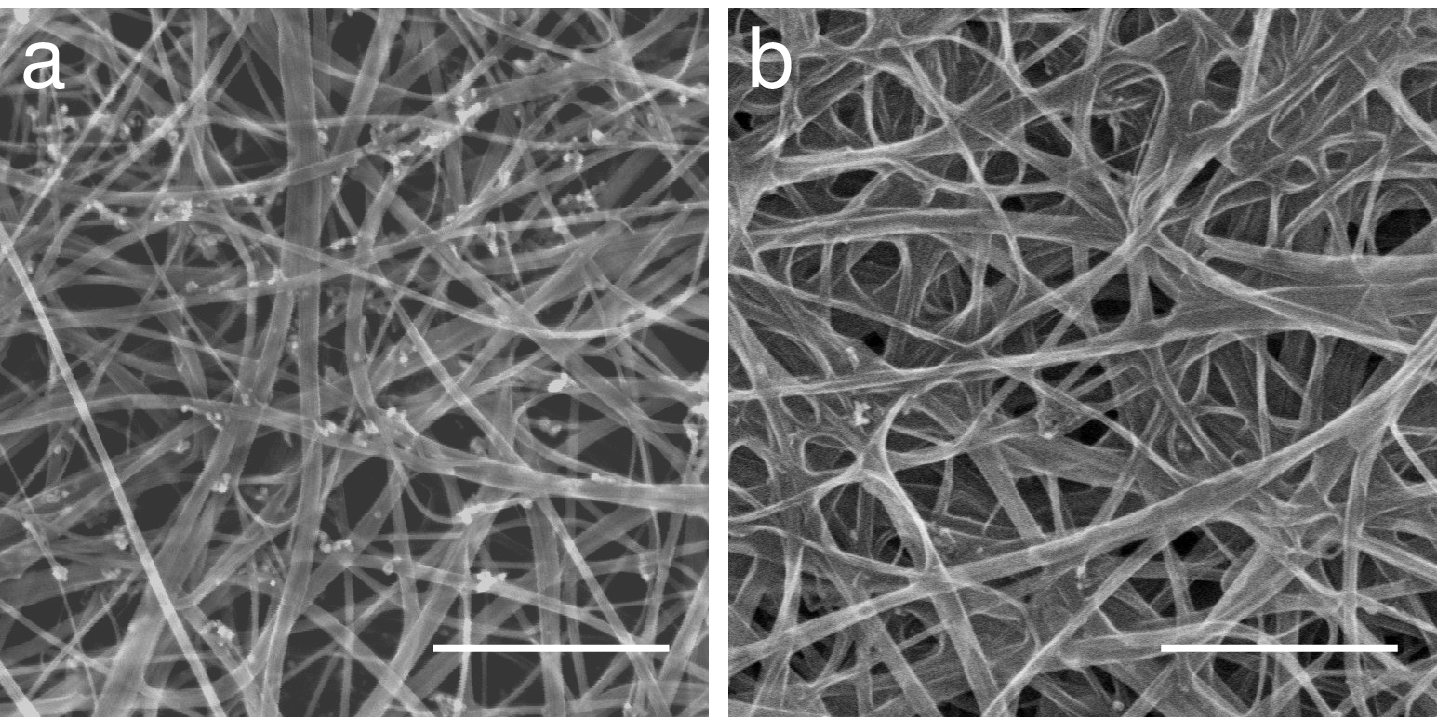}
\caption{\label{fig.densified} Acetone-densified assembly networks of CNT bundles. Scale bars are 500 nm.}
\end{figure}

As discussed in the main text, the entanglement was not altered after liquid densification. Figure \ref{fig.densified}
shows two acetone-densified CNT networks. As compared to Figure \ref{fig.entanglement}a (in the main article), it is
clear that the pore size had decreased from $>$500 nm to $\sim$100--200 nm, while the feature of random distribution and
aggregation structure was maintained. The densification mainly took place along the direction perpendicular to the film
surface, that is, different CNT layers (due to the winding process) were drawn closer, while the shrinkage of film width
and length was less than 2--5\% (but measurable). Therefore, it became possible to introduce resin molecules into the
CNT films without aggregating the CNT bundles.

\begin{figure}[!t]
\centering
\includegraphics[width=.48\textwidth]{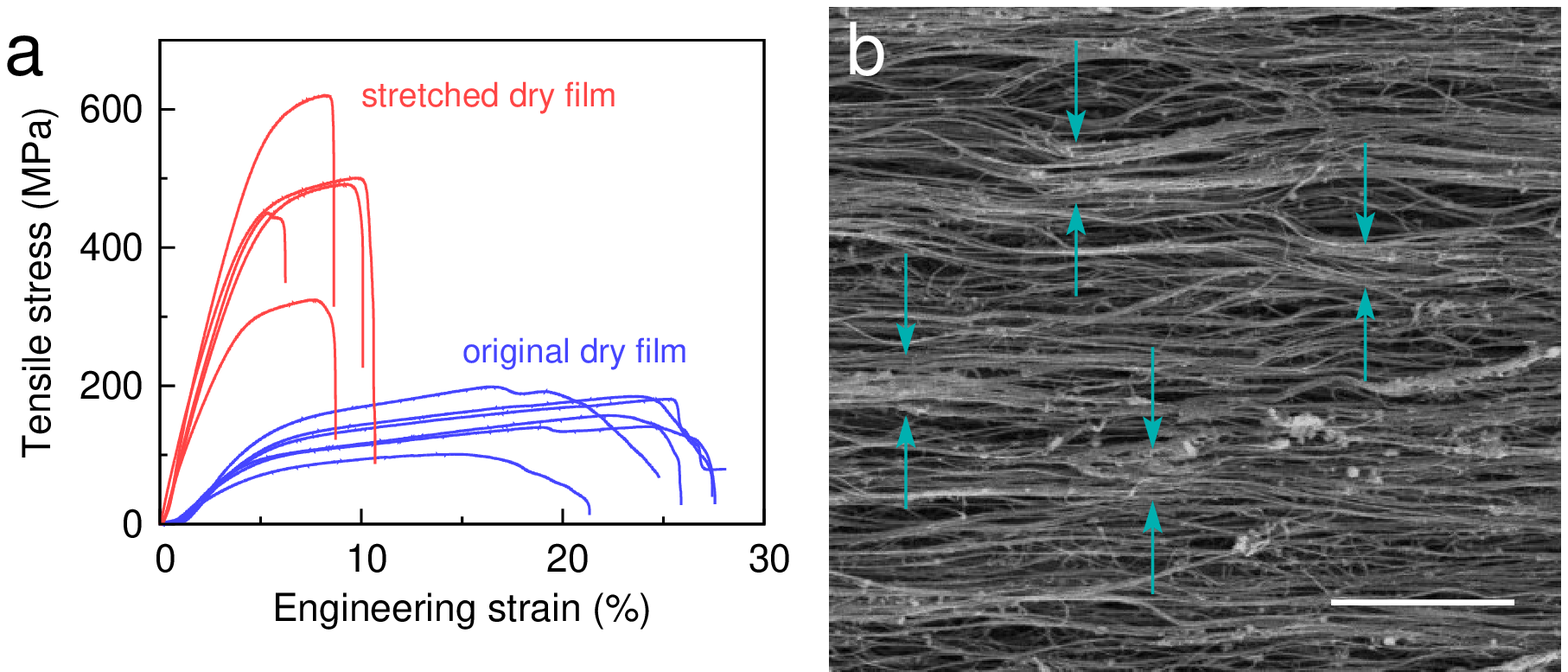}
\caption{\label{fig.raw-tensile} Mechanical properties (a) and CNT alignment (b) of directly stretched films. Scale bar
is 2 \si{\micro}m.}
\end{figure}

\subsection*{3. Mechanical properties of dry films}

As the bismaleimide (BMI) resins had a softening point smaller than 60 \si{\celsius} and demonstrated a viscous liquid
property, we call the resin-impregnated CNT films as ``wet'' films while the pure CNT films as dry ones. The tensile
strength, strain at break, and toughness (the area covered under stress-strain curve) of the as-produced dry films were
up to 180--198 MPa, 20--26\%, and 32--37 MPa, respectively (Figure \ref{fig.raw-tensile}a). Stretching the dry films to
align CNTs is the simplest method to improve the mechanical properties. By stretching directly by 20\% within one minute
and then maintaining the stretching to relax for 10--30 min, the films exhibited improved strengths and toughness, up to
492--620 MPa and 35--40 MPa, respectively. In the stretched film, CNTs became much more aligned (Figure
\ref{fig.raw-tensile}b), making the modulus larger than 12 GPa.

It is important to notice that, during the stretching process the CNTs not only became aligned, but also aggregated to
form large-size bundles (as labelled by arrows in Figure \ref{fig.raw-tensile}b). The aggregation can benefit the load
transfer between CNTs in the stretched dry films, however, might become weak part in composite films by impregnating BMI
resins and then curing them. This is because the aggregated CNTs transfer loads just depending on the intertube sliding
friction while the BMI network can increase significantly the interfacial interactions with CNT surfaces and connect
non-neighboring CNTs, resulting in much more efficient load transfer within polymer-bridged CNTs
\citesupp{s-cheng.qf:2009, s-cheng.qf:2010, s-li.s:2012, s-wang.x:20131}. This means, getting rid of CNT aggregation has
become a challenge as important as improving interfacial interaction and CNT alignment. To further confirm the effect of
aggregation, BMI resins were impregnated into the stretched dry films. As the aggregation existed, such CNT/BMI
composite film exhibited a tensile strength up to only 585--801 MPa, after being cured.

\subsection*{4. High plasticity of ``wet'' films}

\begin{figure}[!t]
\centering
\includegraphics[width=.36\textwidth]{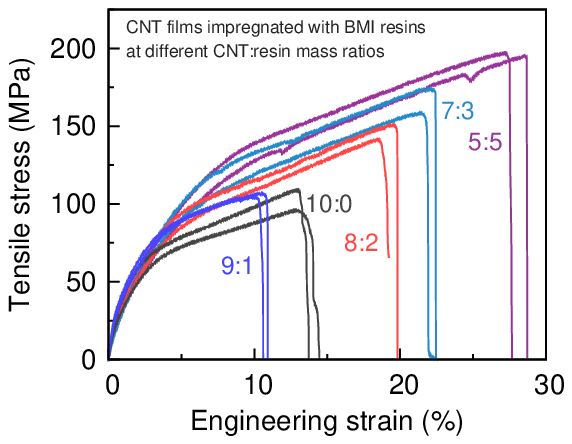}
\caption{\label{fig.wet-tensile} Tensile characterization of CNT films impregnated with different amount of BMI resins.}
\end{figure}

To avoid CNT aggregation, the as-produced CNT films were impregnated with 1 wt\% BMI resin/acetone solutions prior to
the stretching process. By controlling the total volume of solution, ``wet'' films at different CNT-to-resin mass ratios
were produced, namely 9:1, 8:2, 7:3, and 5:5 (similarly, the ratio for dry films was 10:0). The ``wet'' films had
different strain at break at different mass ratios (Figure \ref{fig.wet-tensile}). When a small amount of resins were
impregnated, the resin molecules could not sufficiently cover the CNT surfaces and thus caused significant inhomogeneity
within the film. Therefore the film became easier to fracture. When more resins were used, the film became more and more
plastic and even can be stretched by $\sim$30\%. This means that the wet environment gives the film enhanced
processability.

Notice that the 10:0 sample had a different strain at break from the tensile tests shown in Figure
\ref{fig.raw-tensile}a. This is because the widths of the test samples were 10 and 2 mm for the samples in Figure
\ref{fig.wet-tensile} and \ref{fig.raw-tensile}, respectively. As compared to small samples, assembly defects (in the
as-produced CNT network) make the fracture easier for large ones. In fact, the tensile results of large samples
reflected the real processability, as the resin impregnation and stretching processes were performed on CNT films with a
width of 1 cm.

\subsection*{5. Low-softening-point bismaleimide resins}

\begin{figure}[!t]
\centering
\includegraphics[width=.40\textwidth]{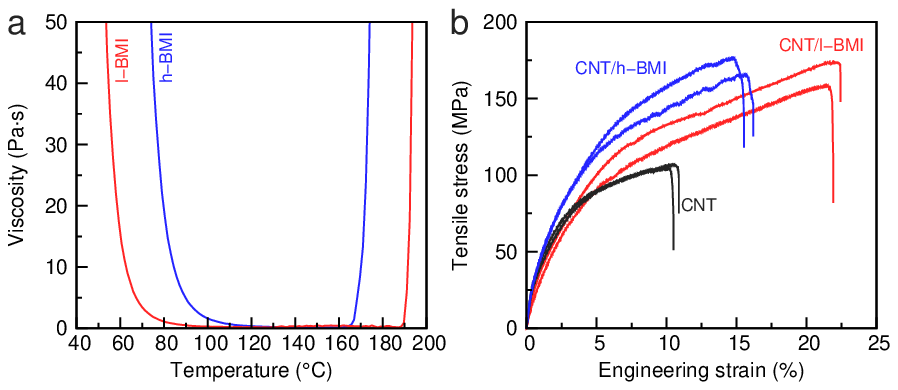}
\caption{\label{fig.BMI} Viscosity of l-BMI and h-BMI resins (a) and their influences on processability of CNT/BMI
composite ``wet'' films (b).}
\end{figure}

The low-softening-point BMI (l-BMI) resins were synthesized by modifying traditional BMI monomers with diallyl bisphenol
A (DBA) \citesupp{s-li.zm:2001}. Besides the DBA modification, the original molecular structure of BMI monomers also
determines remarkably on the softening dynamics. As a comparison study, the DBA-modified resins with larger molecular
weight of monomers, higher molecular rigidity, and thus higher softening point, the h-BMI resins, were also used to show
their influences on the processability of CNT/BMI composite films. Rheological measurement, the viscosity-temperature
relationship, reveals that the viscosity of l-BMI decreased to below 20 \si{\pascal\second} at $\sim$58 \si{\celsius}
while such temperature was $\sim$80 \si{\celsius} for h-BMI (Figure \ref{fig.BMI}a). By impregnating them into CNT films
at the same mass ratio of 7:3, the tensile measurements showed different increase in strain at break. The CNT/h-BMI
``wet'' film fractured at $\sim$15\% while the CNT/l-BMI at more than 22\% (Figure \ref{fig.BMI}b, sample width 10 mm).

\subsection*{6. One-step and multi-step stretching methods}

The stretching methods are schematically shown in Figure \ref{fig.tensile} in the main text. Here we provide more
details about the difference between the one-step and multi-step treatments.

The as-produced CNT films are first impregnated by BMI resin/acetone solution. As acetone has a high infiltration
ability into CNT assemblies, the impregnation can easily make all the CNT bundles surrounded by the resin molecules and
avoid the aggregation of CNTs. ``Wet'' films containing a designed CNT-to-resin mass ratio are prepared by using the
low-softening-point resins. These ``wet'' films can be directly stretched by $\sim$20\% in a one-step way. The one-step
stretching can effectively align the CNTs, however, as the lack of relaxation during the stretching, the entangled or
cross-linked segments can not be fully aligned and thus hinder the densification process.

A multi-step stretching method is used to introduce the relaxation process. In this way, the ``wet'' film is always
stretched slightly in each step, by 2--3\%, and there are 5--10 minutes to relax the film before the next stretching. If
the stretching magnitude $\xi$ depends on the current film length, after $n$ steps, the total stretching magnitude can
be calculated by $(1+\xi)^n-1$. (The stretching magnitude can be confirmed by the final length increase as compared to
the original length.) If the stretching length $x$ is always fixed, the total stretching magnitude is then $nx/L_0$
where $L_0$ is the original film length.

The relaxation process plays important roles in optimizing CNT alignment and level of densification for the ``wet''
films, which can be represented by the mechanical properties of the cured samples and the mass density as well.

\subsection*{7. Mechanical properties of one-step stretched films after curing at mass ratio of 7:3}

The ``wet'' films were cured by a hot-pressing process with a pressure of 6--8 MPa. The curing profile was 140
\si{\celsius} for 0.5 h, 170 \si{\celsius} for 3 h, 220 \si{\celsius} for 2 h, and 250 \si{\celsius} for 3 h. The
introduction of 140 \si{\celsius} treatment was found to be also important to liquidize the resin molecules and further
improved homogeneity in the composite structure.

\begin{figure}[!t]
\centering
\includegraphics[width=.36\textwidth]{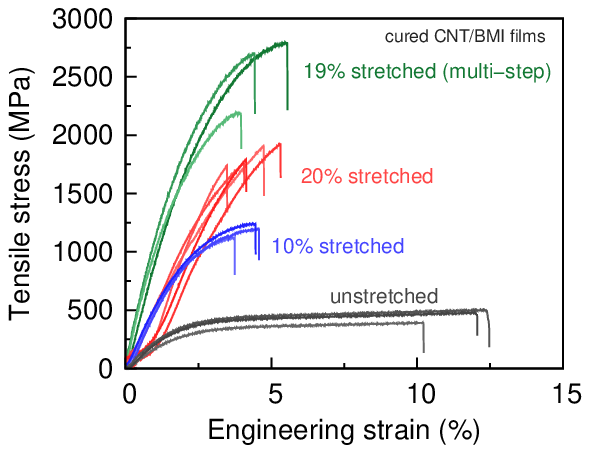}
\caption{\label{fig.one-step} Comparison of tensile properties between different CNT/BMI films which include the
unstretched, one-step stretched (by 10\% and 20\%), and multi-step stretched (by 19\%) films. The CNT-to-resin mass
ratio was 7:3.}
\end{figure}

The one-step stretching was applied by 10\% and 20\%, respectively, on ``wet'' films with the CNT-to-resin mass ratio of
7:3. After being cured, the two stretched composite films exhibited tensile strengths of 1.12--1.24 GPa and 1.74--1.92
GPa, respectively, while the unstretched composite films had strengths less than 500 MPa (Figure \ref{fig.one-step}).
This can be ascribed to the insufficient alignment of CNTs. Furthermore, as compared to the multi-step stretching, the
lack of relaxation process resulted in inhomogeneity in the composite structure. For example, when the ``wet'' films
were stretched by 19\% in the multi-step way, there should not be a significant difference in alignment as compared to
the 20\% one-step stretched films. However, the inhomogeneity in one-step stretched films reduced the efficiency of load
transfer between CNT bundles and thus hindered the increase in elastic modulus. As a result, their tensile strengths of
2.19--2.79 GPa were much larger, at least by 450--870 MPa.

\subsection*{8. Effect of multi-step stretching}

In the multi-step stretching, we set $\xi=3\%$ and the relaxation time to be 10 min, and thus the total stretching
magnitude was $1.03^n-1$. The maximum number of steps differed from sample to sample. For the 7:3 ``wet'' film
$n\leqslant10$ while for the 8:2 film $n\leqslant8$. We also tried to stretch the film by a fixed length of 1 mm in each
step (the original length and width was about 35 mm and 10 mm, respectively). However, we did not observe any difference
between these two treatments. To further increase the maximum stretching magnitude, studies should be performed on the
CVD process where the CNT structure and formation of CNT network are determined.

\begin{figure}[!t]
\centering
\includegraphics[width=.36\textwidth]{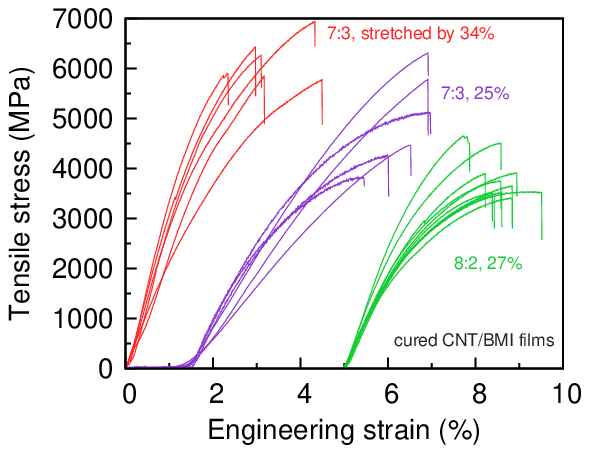}
\caption{\label{fig.multi-step} Stress-stain curves for multi-step stretched CNT/BMI films where the stretching
magnitudes were larger than 25\%. Some plots are shifted for a better comparison.}
\end{figure}

Figure \ref{fig.multi-step} shows the stress-strain curves of cured CNT/BMI films after being multi-step stretched. The
highest performance was found for the 7:3 films being stretched by 34\% ($1.03^{10}-1$). The tensile strengths and
moduli (both in units of GPa) of these tests were 6.940/284.2, 6.438/314.9, 6.265/299.0, 5.907/350.6, 5.842/246.8, and
5.773/211.9, respectively. Even just being stretched up to 25\%, the tensile strength was still found to range from 3.83
to 6.31 GPa, higher than traditional carbon fiber/epoxy composites and recently reported high performance CNT/BMI
composites \citesupp{s-cheng.qf:2009, s-cheng.qf:2010, s-wang.x:20131}.

As compared to the 7:3 mass ratio, the 8:2 ratio resulted in a maximum stretching magnitude of $\sim$27\% and final
tensile strengths ranging from 3.42 to 4.65 GPa. This means that for high performance composites based on polymer
impregnation into CNT assemblies, the polymer content should be around 30 wt\%. When more BMI were introduced into the
film, like mass ratios of 6:4 and 5:5, the BMI aggregation appeared and became structural defects to make the film more
brittle than the 7:3 or 8:2 ones.

\subsection*{9. Herman's orientation factor}

Herman's orientation factor (HOF) is a commonly used parameter to characterize orientation and has been successfully
used for determining the alignment level for CNT arrays \citesupp{s-xu.m:2012}. It takes the value 1 for a system with
perfect orientation parallel to a reference direction, and zero for completely nonoriented samples. The calculation of
HOF is based on the intensity profile $I$ of orientation angle $\phi$ between the structural unit vector and the
reference direction, according to its definition 
\begin{equation}
\text{HOF} \equiv\frac{1}{2}(3\left<\cos^2\phi\right>-1),
\end{equation}
where
\begin{equation}
 \left<\cos^2\phi\right> = \frac{\int_0^{\pi/2}I(\phi)\cos^2\phi \sin\phi \; d\phi}
                                {\int_0^{\pi/2}I(\phi)           \sin\phi \; d\phi}.
\end{equation}
Here the intensity profile was obtained by using the `{\tt Orientation}' property of `{\tt regionprops}' in Matlab,
where an SEM image should be first converted to be strongly contrasted.

\begin{figure}[!t]
\centering
\includegraphics[width=.40\textwidth]{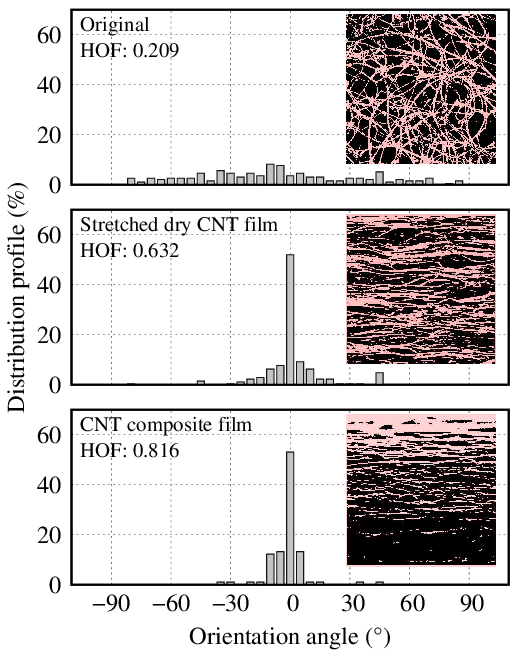}\\
\caption{\label{fig.hof} Intensity profiles of orientation angle for the as-produced CNT film, directly stretched CNT
film, and the super-strong CNT/BMI composite film, whose SEM images are provided in Figure \ref{fig.entanglement}a,
Figure \ref{fig.raw-tensile}b, and Figure \ref{fig.sem}c, respectively.}
\end{figure}

Figure \ref{fig.hof} shows the different intensity profiles for the as-produced CNT film, directly stretched CNT film,
and the super-strong CNT/BMI composite film, respectively. The reference direction was taken to be the horizontal line
for each image. For the original film, the orientation within [-5\si{\degree},5\si{\degree}] was only 9.2\% in the total
counting of the orientation angles. After being stretched, the [-5\si{\degree},5\si{\degree}] fraction increased up to
63.0\% and 74.5\%. However, as the stretching magnitude for the one-step stretching was only 20\% in length, there were
still some orientation angles larger than 40\si{\degree} (6.6\% in the angle counting). According to these profiles, the
HOFs were 0.209, 0.632, and 0.816, respectively. Obviously, the CNT alignment was significantly improved after the
one-step and multi-step stretching treatments.

\subsection*{10. Fracture morphology}

\begin{figure}[!t]
\centering
\includegraphics[width=.40\textwidth]{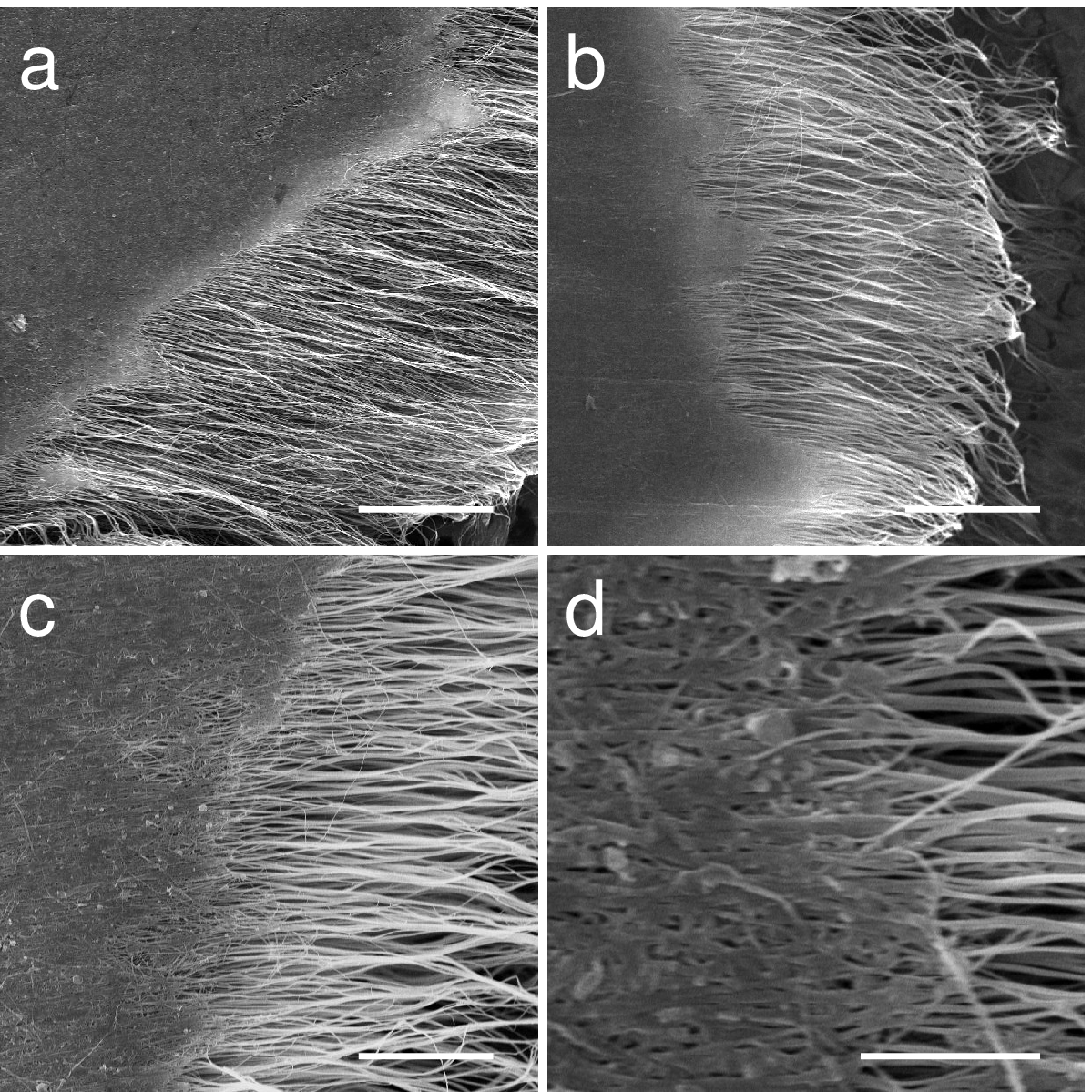}
\caption{\label{fig.fracture} Fracture morphologies of CNT/BMI composite films. Scale bars are 10, 10, 5, and 1
\si{\micro m}, respectively.}
\end{figure}

\begin{figure}[!t]
\centering
\includegraphics[width=.38\textwidth]{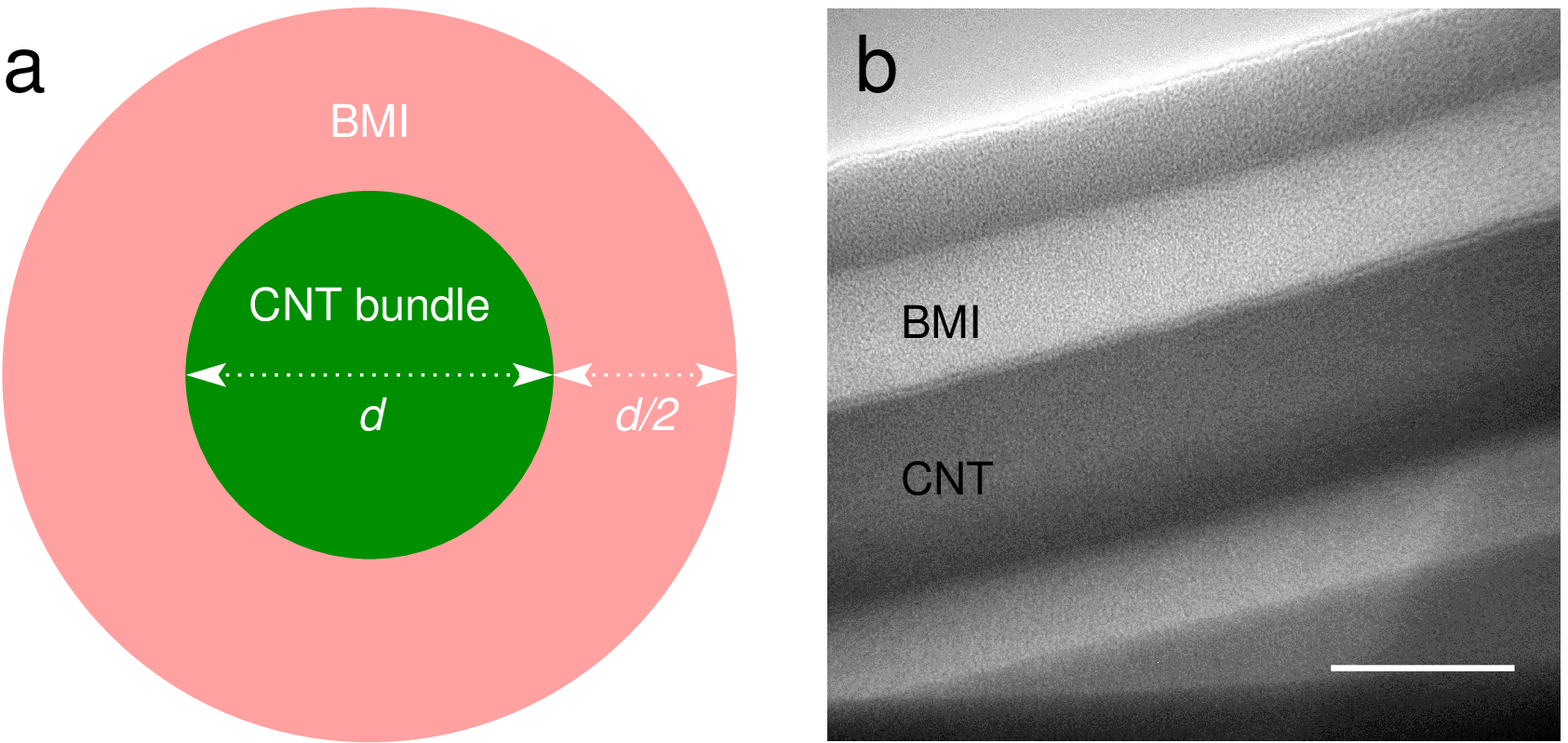}
\caption{\label{fig.bundle} Schematic (a) and TEM characterization (b) of BMI-surrounded CNT bundles. (a) The bundle
width is $d$ and the BMI thickness is $d/2$. The shear contact area is proportional to $\pi d$ and the total area is
$\pi d^2$. (b) The real BMI thickness was in the same order with but slight smaller than the bundle size. Scale bar is
50 nm.}
\end{figure}

Figure \ref{fig.fracture} shows fracture morphologies of CNT/BMI composite films (mass ratio 7:3). Pull-out mechanism
dominated at the fracture and the length of bare CNTs (more than 20 \si{\micro\m}) can be used to estimate the tensile
strength by $\sigma l /d$ (see the schematic shown in Figure \ref{fig.bundle}a), $\sigma$, $d$, and $l$ being the
interfacial shear strength between CNT and BMI, diameter of CNT bundle, and bare length at fracture. Here we assumed
that the CNT bundles were separated from each other by a distance close to the bundle size, and the space between them
was filled by BMI polymers. By using $\sim$30 MPa, $d=50$ nm, and $l=20$ \si{\micro m}, the strength was estimated to be
12 GPa, in nice agreement with experimental measurements.

TEM characterization confirmed that the BMI thickness was in the same order of magnitude with the bundle size (Figure
\ref{fig.bundle}b). According to the strength estimation of $\sigma l /d$, to further improve the mechanical
performance, the interfacial enhancement and reduction in bundle size will be the two major solutions.

Furthermore, from Figure \ref{fig.fracture}d one can find that the CNT bundles were clearly unaggregated and the BMI
polymers did not aggregate but uniformly distributed between the bundles.

\subsection*{11. Film thickness and specific tensile strength}

The calculation of tensile strength requires the information of film width and thickness. The thickness of as-produced
dry films was controlled to be within 10--30 \si{\micro\m}. As the impregnated BMI resins occupied the pores in film,
there was no thickness increase for the ``wet'' films. After being stretched, one-step or multi-step, the thickness
decreased significantly with the stretching magnitude.

\begin{figure}[!t]
\centering
\includegraphics[width=.40\textwidth]{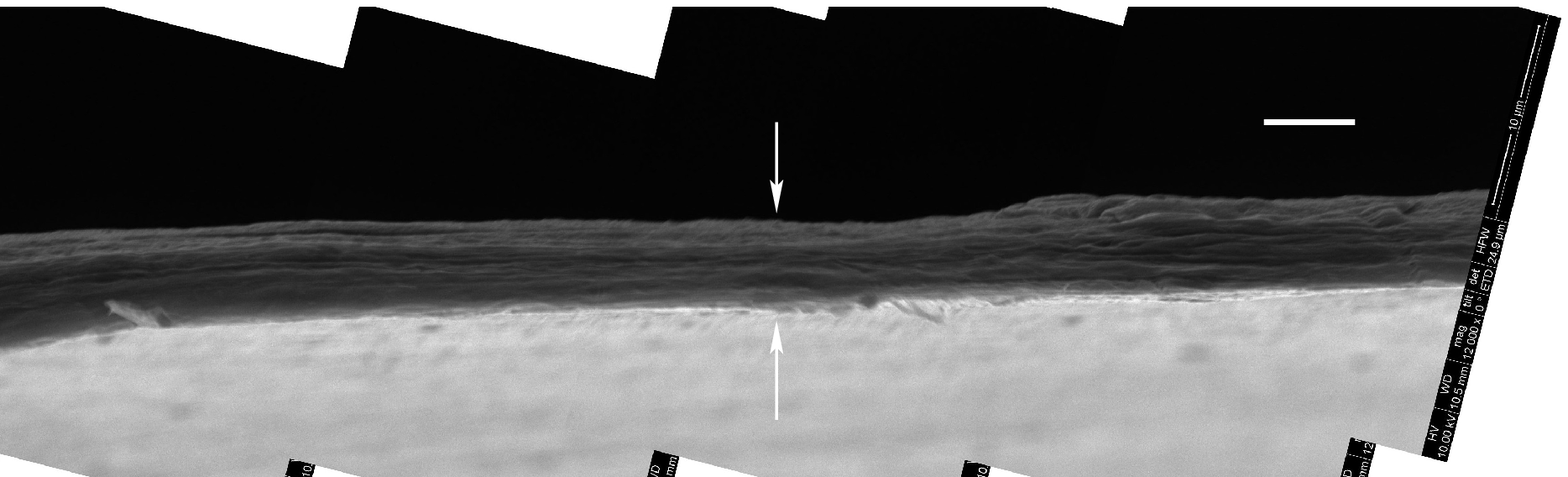}\\
\includegraphics[width=.45\textwidth]{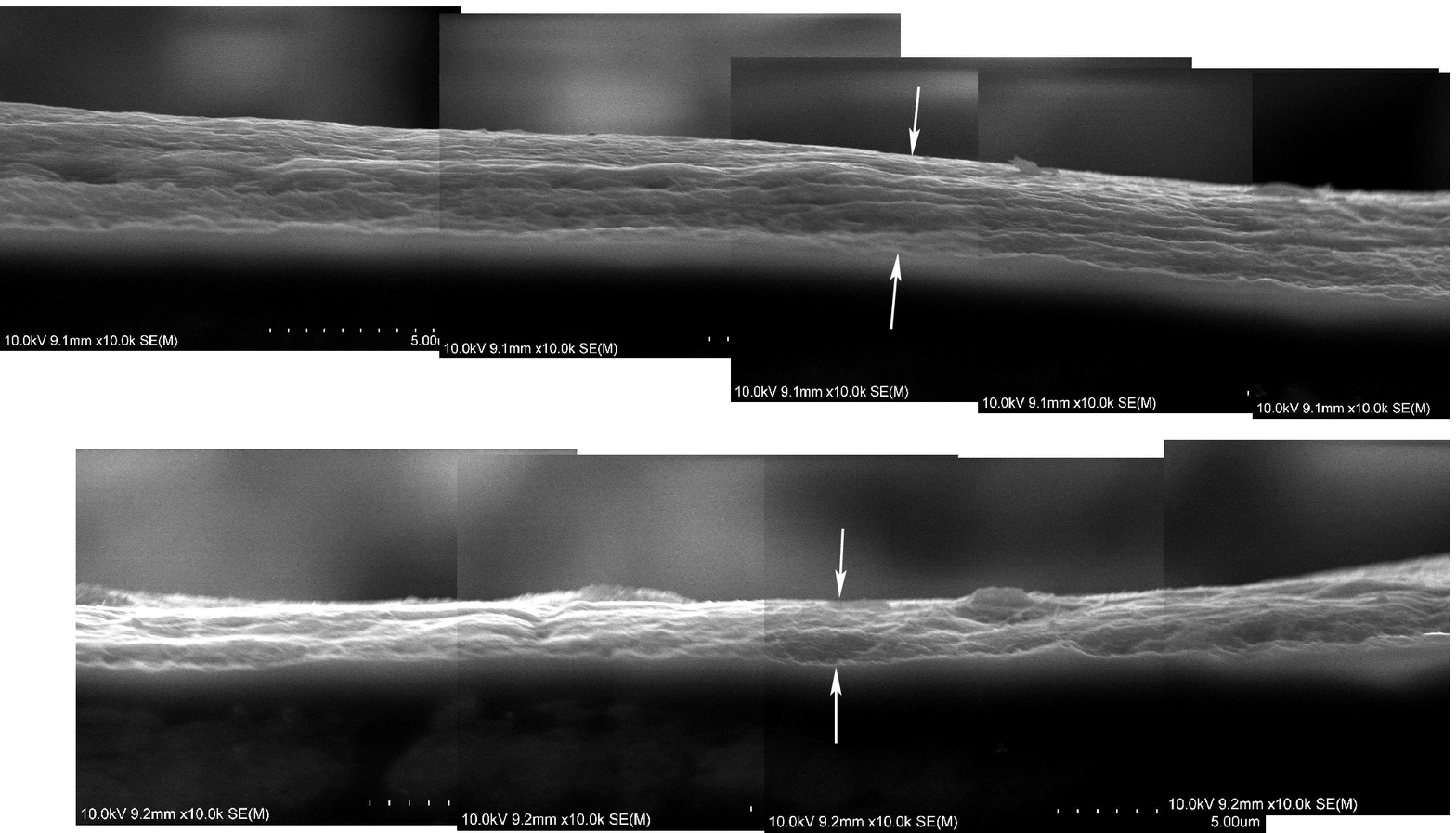}\\
\includegraphics[width=.45\textwidth]{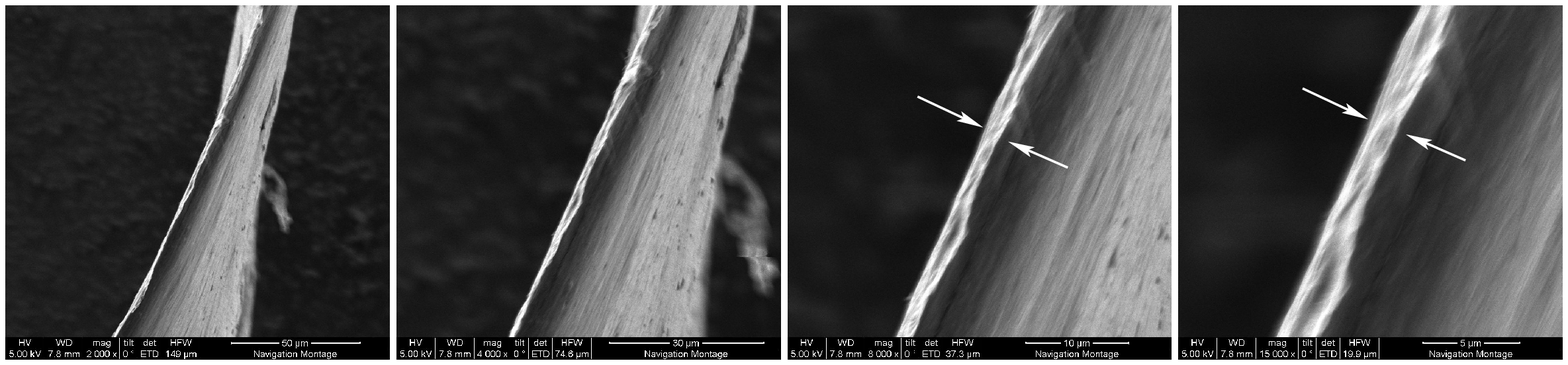}\\
\caption{\label{fig.thickness} SEM characterization of film thickness. The unlabelled scale bars are all 5
\si{\micro\m}.}
\end{figure}

Figure \ref{fig.thickness} shows thicknesses of four stretched and cured composite films, with stretching magnitude
ranged from $\sim$20\% to $\sim$34\% (from up to bottom). The final thicknesses was measured to be 5, 2.8, 2, and 1.7
\si{\micro\m}, respectively. To avoid overestimation, we did not use the thickness of the thinnest segment to calculate
the tensile strength, but the measurement on the thickest segment. For example, the thickness of the 34\% stretched film
was 2.5--3 \si{\micro\m}.

\begin{figure}[!t]
\centering
\includegraphics[width=.45\textwidth]{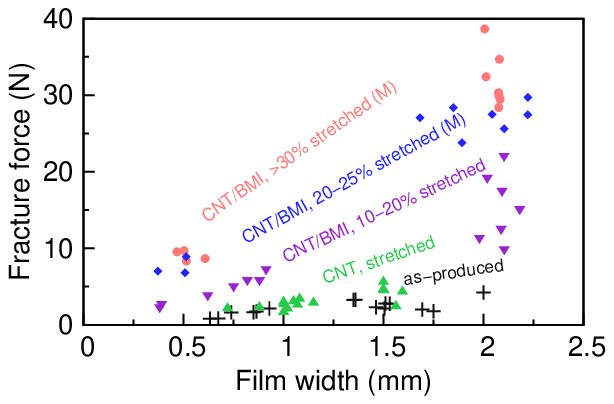}
\caption{\label{fig.load} Fracture load as a function of sample width for different films.}
\end{figure}

As discussed in the main text, one can calculate the specific strength by measuring the fracture force and mass density
of the CNT/BMI composite films. For the multi-step stretched films by $\sim$25\%, the total mass for a 2 cm$\times$1 cm
sample was 1.15 mg, corresponding to an area density of 0.58 mg/cm$^2$. The total force to fracture such film was about
16.5 N/mm in average (Figure \ref{fig.load}), and thus the specific strength (by dividing the force per width by the
area density) was $\sim$2.87 N/tex. When the stretching magnitude was improved to 34\%, the fracture force per width was
$\sim$19.5 N/mm, the area density decreased slightly to 0.46 mg/cm$^2$, and the specific strength was $\sim$4.24 N/tex.
As the mass density was about 1.55 g/cc, the product of specific strength and volumetric density was 6.57 GPa. This
means that all the measured data, including the force, film thickness, and mass density were completely consistent.

\subsection*{12. Summary}

We provide here detailed information of CNT growth, BMI resins, and impregnation and stretching techniques. Based on the
unique properties of raw materials (entanglement, unaggregation, high plasticity, and low softening point) and the
multi-step stretching method which was applied on resin-impregnated CNT films, we have been able to obtain a magic
composite structure where neither CNTs nor polymers formed aggregated phases, a big step to approach the ideal composite
structure to fully utilize all the CNT surfaces in load transferring. The highest tensile strength was 6.94 GPa (or 4.24
N/tex), much higher than the strength of carbon fiber reinforced polymers. The CNT/BMI composite films also exhibited
high ability to conduct electricity.


\begin{thebibliography}{28}%
\bibitem{ha.tlb:2013}
Ha, T. L.~B., Quan, T.~M., Vu, D.~N. \& Si, D.~M.
Naturally Derived Biomaterials: Preparation and Application, in:
{\em Regenerative Medicine and Tissue Engineering} (ed. Andrades, J.~A.)
  Ch.~11,  pp 247--274.
\newblock (InTech - Open Access, Rijeka, Croatia 2013).

\bibitem{giesa.t:2011}
Giesa, T., Arslan, M., Pugno, N.~M. \& Buehler, M.~J.
Nanoconfinement of Spider Silk Fibrils Begets Superior Strength, Extensibility, and Toughness.
\newblock {\em Nano Lett.}{ \bf 11}, 5038--5046 (2011).

\bibitem{preston.cm:1992}
Preston, C.~M. \& Sayer, B.~G.
What's in a nutshell: an investigation of structure by carbon-13 cross-polarization magic-angle spinning nuclear magnetic resonance spectroscopy.
\newblock {\em J. Agric. Food Chem.}{ \bf 40}, 206--210 (1992).

\bibitem{porter.sm:2007}
Porter, S.~M.
Seawater Chemistry and Early Carbonate Biomineralization.
\newblock {\em Science}{ \bf 316}, 1302--1302 (2007).

\bibitem{cheng.qf:2014}
Cheng, Q., Jiang, L. \& Tang, Z.
Bioinspired Layered Materials with Superior Mechanical Performance.
\newblock {\em Acc. Chem. Res.}{ \bf 47}, 1256--1266 (2014).

\bibitem{baughman.rh:2002}
Baughman, R.~H., Zakhidov, A.~A. \& de Heer, W.~A.
Carbon Nanotubes -- the Route Toward Applications.
\newblock {\em Science}{ \bf 297}, 787--792 (2002).

\bibitem{coleman.jn:2006}
Coleman, J.~N., Khan, U., Blau, W.~J. \& Gun'ko, Y.~K.
Small but strong: A review of the mechanical properties of carbon nanotube-polymer composites.
\newblock {\em Carbon}{ \bf 44}, 1624--1652 (2006).

\bibitem{moniruzzaman.m:2006}
Moniruzzaman, M. \& Winey, K.~I.
Polymer Nanocomposites Containing Carbon Nanotubes.
\newblock {\em Macromolecules}{ \bf 39}, 5194--5205 (2006).

\bibitem{liu.lq:2011}
Liu, L., Ma, W. \& Zhang, Z.
Macroscopic Carbon Nanotube Assemblies: Preparation, Properties, and Potential Applications.
\newblock {\em Small}{ \bf 7}, 1504--1520 (2011).

\bibitem{kong.lr:2014}
Kong, L. \& Chen, W.
Carbon Nanotube and Graphene-based Bioinspired Electrochemical Actuators.
\newblock {\em Adv. Mater.}{ \bf 26}, 1025--1043 (2014).

\bibitem{coleman.jn:20062}
Coleman, J.~N., Khan, U. \& Gun'ko, Y.~K.
Mechanical Reinforcement of Polymers Using Carbon Nanotubes.
\newblock {\em Adv. Mater.}{ \bf 18}, 689--706 (2006).

\bibitem{fiedler.b:2006}
Fiedler, B., Gojny, F.~H., Wichmann, M. H.~G., Nolte, M. C.~M. \& Schulte, K.
Fundamental aspects of nano-reinforced composites.
\newblock {\em Compos. Sci. Technol.}{ \bf 66}, 3115--3125 (2006).

\bibitem{xie.xl:2005}
Xie, X.-L., Mai, Y.-W. \& Zhou, X.-P.
Dispersion and alignment of carbon nanotubes in polymer matrix: A review.
\newblock {\em Mater. Sci. Eng. R}{ \bf 49}, 89--112 (2005).

\bibitem{spitalsky.z:2010}
Spitalsky, Z., Tasis, D., Papagelis, K. \& Galiotis, C.
Carbon nanotube-polymer composites: Chemistry, processing, mechanical and electrical properties.
\newblock {\em Prog. Polym. Sci.}{ \bf 35}, 357--401 (2010).

\bibitem{rahmat.m:2011}
Rahmat, M. \& Hubert, P.
Carbon nanotube-polymer interactions in nanocomposites: A review.
\newblock {\em Compos. Sci. Technol.}{ \bf 72}, 72--84 (2011).

\bibitem{zhang.m:2005}
Zhang, M. {\it et al.}
Strong, Transparent, Multifunctional, Carbon Nanotube Sheets.
\newblock {\em Science}{ \bf 309}, 1215--1219 (2005).

\bibitem{cheng.qf:2009}
Cheng, Q. {\it et al.}
High Mechanical Performance Composite Conductor: Multi-Walled Carbon Nanotube Sheet/Bismaleimide Nanocomposites.
\newblock {\em Adv. Funct. Mater.}{ \bf 19}, 3219--3225 (2009).

\bibitem{cheng.qf:2010}
Cheng, Q., Wang, B., Zhang, C. \& Liang, Z.
Functionalized Carbon-Nanotube Sheet/Bismaleimide Nanocomposites: Mechanical and Electrical Performance Beyond Carbon-Fiber Composites.
\newblock {\em Small}{ \bf 6}, 763--767 (2010).

\bibitem{liu.w:2011}
Liu, W. {\it et al.}
Producing superior composites by winding carbon nanotubes onto a mandrel under a poly(vinyl alcohol) spray.
\newblock {\em Carbon}{ \bf 49}, 4786--4791 (2011).

\bibitem{di.jt:2012}
Di, J. {\it et al.}
Dry-Processable Carbon Nanotubes for Functional Devices and Composites.
\newblock {\em ACS Nano}{ \bf 6}, 5457--5464 (2012).

\bibitem{wang.x:20131}
Wang, X. {\it et al.}
Ultrastrong, Stiff and Multifunctional Carbon Nanotube Composites.
\newblock {\em Mater. Res. Lett.}{ \bf 1}, 19--25 (2013).

\bibitem{t300}
TORAYCA® carbon fibers T300 data sheet: http://www.toraycfa.com/pdfs/T300DataSheet.pdf.

\bibitem{wagner.hd:2007}
Wagner, H.~D.
Nanocomposites: Paving the way to stronger materials.
\newblock {\em Nat. Nanotechnol.}{ \bf 2}, 742--744 (2007).

\bibitem{li.yl:2004}
Li, Y.-L., Kinloch, I.~A. \& Windle, A.~H.
Direct Spinning of Carbon Nanotube Fibers from Chemical Vapor Deposition Synthesis.
\newblock {\em Science}{ \bf 304}, 276--278 (2004).

\bibitem{zhang.m:2004}
Zhang, M., Atkinson, K.~R. \& Baughman, R.~H.
Multifunctional Carbon Nanotube Yarns by Downsizing an Ancient Technology.
\newblock {\em Science}{ \bf 306}, 1358--1361 (2004).

\bibitem{li.zm:2001}
Li, Z., Yang, M., Huang, R., Zhang, M. \& Feng, J.
Bismaleimide resin modified with diallyl bisphenol A and diallyl p-phenyl diamine for resin transfer molding.
\newblock {\em J. Appl. Polym. Sci.}{ \bf 80}, 2245--2250 (2001).

\bibitem{xu.m:2012}
Xu, M., Futaba, D.~N., Yumura, M. \& Hata, K.
Alignment Control of Carbon Nanotube Forest from Random to Nearly Perfectly Aligned by Utilizing the Crowding Effect.
\newblock {\em ACS Nano}{ \bf 6}, 5837--5844 (2012).

\bibitem{xu.m:2010}
Xu, M., Futaba, D.~N., Yamada, T., Yumura, M. \& Hata, K.
Carbon Nanotubes with Temperature-Invariant Viscoelasticity from -196\si{\degree} to 1000\si{\degree}C.
\newblock {\em Science}{ \bf 330}, 1364--1368 (2010).

\end{thebibliography}

\begin{thebibliography}{1}
\bibitem{s-li.yl:2004} Y.-L. Li, I.~A. Kinloch, A.~H. Windle, \emph{Science} \textbf{2004}, \emph{304}, 276.

\bibitem{s-cheng.qf:2009} Q.~Cheng, J.~Bao, J.~Park, Z.~Liang, C.~Zhang, B.~Wang, \emph{Adv. Funct. Mater.} \textbf{2009}, \emph{19}, 3219.

\bibitem{s-cheng.qf:2010} Q.~Cheng, B.~Wang, C.~Zhang, Z.~Liang, \emph{Small} \textbf{2010}, \emph{6}, 763.

\bibitem{s-li.s:2012} S.~Li, X.~Zhang, J.~Zhao, F.~Meng, G.~Xu, Z.~Yong, J.~Jia, Z.~Zhang, Q.~Li, \emph{Compos. Sci. Technol.} \textbf{2012}, \emph{72}, 1402.

\bibitem{s-wang.x:20131} X.~Wang, Z.~Z. Yong, Q.~W. Li, P.~D. Bradford, W.~Liu, D.~S. Tucker, W.~Cai, H.~Wang, F.~G. Yuan, Y.~T. Zhu, \emph{Mater. Res. Lett.} \textbf{2013}, \emph{1}, 19.

\bibitem{s-li.zm:2001} Li, Z., Yang, M., Huang, R., Zhang, M., and Feng, J. \newblock {\em J. Appl. Polym. Sci.}{ \bf 80}(12), 2245--2250 (2001).

\bibitem{s-xu.m:2012} Xu, M., Futaba, D.~N., Yumura, M., and Hata, K. \newblock {\em ACS Nano}{ \bf 6}(7), 5837--5844 (2012).

\end{thebibliography}
\end{document}